# Advancing a Model of Students' Intentional Persistence in Machine Learning and Artificial Intelligence

Sharon Ferguson[1], Katherine Mao[1], James Magarian[2], Alison Olechowski[1]

[1] University of Toronto

[2] Massachusetts Institute of Technology

**Abstract**

Machine Learning and Artificial Intelligence are powering the applications we use, the decisions we make, and the decisions made about us. We have seen numerous examples of non-equitable outcomes, from facial recognition algorithms, recidivism algorithms, and resume reviewing algorithms, when they are designed without diversity in mind. As Machine Learning (ML) and Artificial Intelligence (AI) expand into more areas of our lives, we must take action to promote diversity among those working in this field. A critical step in this work is understanding why some students who choose to study ML/AI later leave the field.

While the persistence of diverse populations has been studied in engineering specifically, and Science, Technology, Engineering and Math (STEM) more generally, there is a lack of research investigating factors that influence persistence in ML/AI. In this work, we present the advancement of a model of intentional persistence in ML/AI in order to identify areas for improvement. We surveyed undergraduate and graduate students enrolled in ML/AI courses at a major North American university in fall 2021. We examine persistence across demographic groups, such as gender, international student status, student loan status, and visible minority status. We investigate independent variables that distinguish ML/AI from existing studies of persistence in STEM, such as the varying emphasis on non-technical skills, the ambiguous ethical implications of the work, and the highly competitive and lucrative nature of the field.

Our findings suggest that short-term intentional persistence in ML/AI is associated with academic enrollment factors such as major and level of study. In terms of long-term intentional persistence, we found that measures of professional role confidence developed to study persistence in engineering are also important predictors of intent to remain in ML/AI. Unique to our study, we show that wanting your work to have a positive social benefit is a negative predictor of long-term intentional persistence in ML/AI, and women generally care more about this. We find some evidence that having high confidence in non-technical interpersonal skills may also be a positive predictor of long-term intentional persistence. We provide recommendations to educators to meaningfully discuss ML/AI ethics in classes and encourage the development of interpersonal skills to help increase diversity in the field.

**Introduction**

Artificial Intelligence (AI) and machine learning (ML) have demonstrated tremendous capacity and promise to revolutionize data analysis and decision-making across sectors, including engineering [1]–[3]. Global ML/AI hiring continues to rise, and the world's top universities have increased their investment in AI education. In fact, from 2017-2021, the number of ML/AI courses at the undergraduate level has increased by 100%, and at the graduate level by 40% [3].

While the shift towards ML/AI is undeniable, we must also recognize that where diversity numbers exist, we see a stark and broad diversity problem in the field, as has long been reported in computer science and STEM [4], [5]. Only 18% of graduates from AI PhD programs are women in North America [3], and 22% globally [6]. Ethnicity and race statistics, when available, show rates of roughly 4% Black representation in the AI PhD graduates and technology work sectors [3], [6]. The emergence of initiatives and organizations such as "Queer In AI," "Black In AI," and "Indigenous AI" indicate that there is a need for support and outreach towards underrepresented groups in this field [3].

This lack of diversity has multiple important consequences. First, lacking a diverse field of workers, ML/AI algorithm designers leveraging their perspectives and experiences may be more susceptible to bias, which can perpetuate societal inequalities. Diverse groups would be more likely to notice the potential for algorithmic bias, such as misclassification (see [7]) and misrepresentative data sets (see [8]). Next, diverse teams have been shown to lead to more inclusive, high quality, and innovative outcomes in many contexts [9]–[11]. Finally, diversity segregation in the job market leaves minorities out of high-paying, influential, and desirable jobs [6].

A complex problem like this requires a multi-dimensional solution. In this paper, we aim to better understand the persistence of post-secondary and graduate STEM students in ML/AI. Specifically, we aim to advance a model of student persistence by including additional variables that we suspect will explain greater variance in the unique case of ML/AI, and this paper sets up and test hypotheses to do this. In other contexts, understanding persistence has been shown to be important for diversity and representation in the workforce [12]–[16]. Yet we see that broad studies of persistence often show mixed results; studies are needed that narrow the scope to specific areas of study. While engineering and computer science have been the focus of studies of persistence and retention [16], [17], the persistence of students in the field of ML/AI has only recently been investigated [18]. This new survey builds on this initial work, investigating additional independent variables that may distinguish ML/AI from existing studies of persistence in STEM, such as the varying emphasis on non-technical skills [19], [20], the ambiguous ethical implications of the work [21]–[23], and the highly competitive and lucrative nature of the field [24]–[26]. We conjecture that these features of ML/AI may result in student persistence being driven in different ways than in traditional engineering or STEM.

In this paper, we first review the literature to identify the theoretically relevant relationships and use these to generate hypotheses. Next, we test our hypotheses with a survey dataset of 159 responses from students taking ML/AI courses at a major North American university. We found that measures of professional role confidence developed to study persistence in engineering are also important predictors of intent to remain in ML/AI. We are also the first to identify a potential driver of the gender gap in ML/AI: wanting your work to have a positive social benefit is a negative predictor of long-term intentional persistence, and women generally care more about this. We also find some evidence that those with high confidence in their non-technical interpersonal skills may be more likely to intend to persist in the field. Lastly, that our respondents' course choices were influenced by the popularity of ML/AI as a subject, suggesting that not all students who take a ML/AI course do so for career preparation purposes. We provide recommendations to educators to emphasize the positive benefits that ML/AI can have on society, while meaningfully discussing the ethical challenges, and to encourage the development of interpersonal skills in these courses.

**Background**

Here we discuss related persistence research and build the theory behind our hypotheses. We begin by explaining our dependent variable of intentional persistence, followed by our hypothesized independent variables, other important independent variables, and conclude with a summary of the relationship between individual characteristics and persistence.

*Intentional Persistence*

Persistence refers to a commitment to remain in a profession or field. The term is often used to examine student persistence in a field of study up to [15], [16], [27]–[29] and after graduation [13], [18], [30]. Scholars have studied the persistence of underrepresented groups, such as women [13], [16], [18] or ethnic minority groups in STEM [31]. Various factors are shown to be associated with persistence, including identity [16], financial aid status [13], [28] [15], confidence [16], [18], and institutional culture [27]; a more in depth discussion of factors important to this study are included in the following sections.

Persistence is commonly measured as part of a longitudinal study where researchers follow students through their university program to track whether they graduate from the program [18], [21], or which programs they switch into [16]. This is referred to as *behavioural persistence* as it measures whether students actually do persist [16]. On the other hand, *intentional persistence* measures whether students intend to remain in the field at some point in the future [16], [18], [30]. In this study, we measure intentional persistence as a first step in understanding persistence in this new and evolving field. Specifically, we measure long-term intentional persistence (five years in the future), as in [16], and introduce a new measure of short-term intentional persistence (within their univerisity program).

*Hypothesized Independent Variables*

To advance our model of student persistence in ML/AI, we develop hypotheses related to three key independent variables: social benefit interest, non-technical skills, and social belonging confidence.

Research in other fields has shown that social benefit interest – feeling like your work has a positive impact on society – is an important factor of persistence and overall job satisfaction [32][33]. Social benefit interest has been studied in terms of gender: women place more importance on altruistic values at work [34]; are more likely to explain their interest in engineering based on societal contribution [12], [35]; are more likely to specialize in "socially conscious" engineering disciplines [36]; and rate impact-driven work as important more often than their peers [24]. High social benefit interest is often studied in relation to public sector work

[32], [37], but has also been investigated in engineering [38]. Although there has been a lot of recent research into ethical AI use [39], researchers have not yet investigated the social work values of ML/AI practitioners.

We can look to engineering as a related field to start building our theory. In general, engineering is believed to be a profession that has a positive impact on the world [40], and many students cite this as a reason for choosing to study it [35]. Despite these stereotypes, Litchfield and Javernick-Will [41] found that socially engaged engineers are misaligned with current engineering careers, and need to find more meaningful work outside of engineering. Engineering is an established field, and most people have a general opinion on the social benefit of pursuing engineering; however, ML/AI is a newly emerging and developing field with split opinions [23], [42]. A recent Pew Research Center study found that the general public in North America and Europe were split on whether the development of AI is good for society, but those in Asia were more likely to view AI positively [23]. Further, a recent Boston Consulting Group study found that many STEM students view data science and ML/AI careers as low-impact and low purpose [24]. Numerous studies have found that men express a more positive view of AI [22], [23], [42]. In contrast, women are more concerned about personal data collection with AI tools and believe that AI will result in less human interaction [22]. In fact, Hoffman et al. in 2018 suggested that we have to emphasize the meaningful nature of work in Computer Science and ML/AI in order to retain more women [21]. Given these findings, we hypothesize that women will show higher levels of social benefit interest, and that the divided opinion of the impact of AI on society means that social benefit interest will be negatively associated with long-term intentional persistence.

<u>Hypothesis 1A: Among students, women will have higher levels of social benefit interest</u>
<u>Hypothesis 1B: Social benefit interest will negatively predict long-term intentional persistence</u>

Technical confidence is the self-assessment of one's abilities or self-efficacy in specific skills related to a field. Historically, a major deterrent barring women from STEM fields was lower levels of math self-assessment due to negative societal stereotypes [43], even among those pursuing degrees and working in careers where high mathematical competency is required [16], [44]–[46]. Further, women begin to lose confidence in their scientific abilities after their first year of university, without a change in measured performance [47]. Low technical confidence has been shown to affect entry and persistence in early STEM education [44], [46] as well as entry level salary expectations [48]. However, recent studies are revealing that math self-confidence is becoming less important for persistence [4], [44], with some studies showing no influence on persistence within students already pursuing STEM degrees [16], [18]. While there are a vast number of technical skills required for ML/AI jobs, they generally cluster around math and statistics or specific computer programming skills [19], [20], [49], [50]. In this study, we measure technical confidence as part of the expertise confidence measure discussed in the next section. However, a larger focus in this study is non-technical self-assessment.

Some STEM persistence research argues that non-technical skills are important predictors of persistence [51]. Specifically, researchers found that holding math and science skills constant, those who had stronger reading and writing abilities were less likely to pursue STEM subjects [51], [52]. In engineering, the Canadian and American accreditation boards stress the importance of building non-technical skills in students; notably, communicating complex engineering concepts (through reading, writing, speaking, and listening), and identifying and formulating engineering problems [53], [54].

Although women are often stereotyped as being stronger in non-technical skills [12], some studies have shown no difference in confidence levels in professional and interpersonal skills between women and men [55]. These gendered stereotypes have important consequences; studies have shown that non-technical or "soft" skills are often used as support for promoting women into management positions and out of technical roles, furthering the negative stereotypes related to women's technical skills [56]. This same pattern has been reported in engineering student teams [12]. Due to these stereotypes, and research that has shown women have higher reading/writing scores [51], we hypothesize that women will report higher levels of non-technical self-assessment.

To understand how salient non-technical skills are to career choice in ML/AI specifically, we start by looking at engineering. Past work has shown that engineering students who anticipate a promotion to a formal leadership position by the age of 25 often expect to work in fields outside of engineering, such as project or product management, technical or management consulting, finance, or venture capital [30]. This suggests that students who believe that they have strong leadership skills and enjoy leadership positions intend to leave engineering. Research has also shown that "soft skill" courses in computer engineering programs have lower satisfaction ratings, yet students report a positive attitude toward learning them [57]. Many studies have investigated the technical and non-technical skills required for ML/AI roles by scraping job postings. The non-technical skills reported include communication [19], [20], [49], [50], employee attitude [19], [50], time management [19], teamwork [19], [20], [49], [50], leadership [20], [49], [50], business acumen [49], [50], and project management [20], [49], [50]. We chose to narrow the scope of skills in this paper to interpersonal skills, and we conceptualize non-technical self-assessment as the combination of communication skills, teamwork skills, and leadership skills. Research shows that non-technical skills are important for ML/AI positions, but students may not yet be aware that their "softer" skills are beneficial in these roles. For this reason, we hypothesize that high levels of non-technical self-assessment will be negatively associated with short- and long-term intentional persistence.

<u>Hypothesis 2A: Among students, women will have higher levels of non-technical self-assessment</u>
<u>Hypothesis 2B: Non-technical self-assessment will negatively predict short-term intentional persistence</u>
<u>Hypothesis 2C: Non-technical self-assessment will negatively predict long-term intentional persistence</u>

Social belonging confidence is the degree to which a person feels that they will fit in with the social and cultural aspects of a profession and develop meaningful relationships with their peers. We investigate social belonging through the lens of confidence because it is an internal sense of how much one belongs in a social setting. Further, belonging uncertainty, as defined by Walton and Cohen [58], can be interpreted as a lack of this confidence.

A sense of belonging has been shown to be a positive predictor for career interest [59], [60] and intentional persistence for students in STEM [61]–[63]. Belonging uncertainty is closely tied to stereotype threat and discrimination, thus, marginalized groups may experience lower confidence in social belonging [64]. For instance, it has been shown that female STEM students report lower average levels of social belonging confidence compared to male students [63], [65]–[67] and that belonging confidence is a stronger predictor for persistence in women compared to men [59], [62]. Further, some studies suggest that social belonging confidence was a stronger predictor of persistence than technical confidence [61], [65]. This solidifies the need to study the relationship between social belonging confidence and persistence in the field of ML/AI, and motivates the following two hypotheses.

Hypothesis 3A: Among students, men will have higher levels of social belonging confidence
Hypothesis 3B: Social belonging confidence will positively predict long-term intentional persistence

*Other Key Independent Variables*

While we formally hypothesize the relationship with three key variables and persistence, we include an additional seven variables that are reported to be related to persistence in past work, or are believed to distinguish ML/AI from previous work in engineering and STEM. These variables include individual measures of professional role confidence (comprised of expertise confidence and career-fit confidence), reported measures of the toxicity of the ML/AI environment, early exposure to ML/AI, and measures related to the societal view of ML/AI as a lucrative and in-demand career.

Professional role confidence is the degree to which a person feels confident in their competence in, satisfaction with, and identity within a profession of interest [16]. This is broken down into two dimensions, expertise and career-fit confidence, as defined by Cech et al. [16]. Expertise confidence is a holistic view of one's confidence in having the competency in skills and knowledge required to succeed in a profession [16]. Career-fit confidence refers to the degree to which the day-to-day work and values of a profession align with the interests and beliefs of an individual [16]. It has been shown that both dimensions of professional role confidence are strong positive predictors of persistence of students in engineering [16] and in ML/AI [18].

Discrimination against women and racial minorities in STEM has been documented [68], [69] and continues in today's workplaces. According to a survey conducted by the PEW Research Centre in 2017, 64% of women in computer occupations have experienced discrimination, compared to 16% of men, and 62% of Black STEM professionals have experienced race-based discrimination, compared to 13% of white STEM professionals [70]. In educational settings, those who have negative experiences with peers and instructors are less likely to be committed to engineering [71]. Further, experiencing discrimination during university has been shown to be negatively associated with self-efficacy and persistence in STEM for women [62], [47] especially if the discrimination was perpetrated by a faculty member [18], [72]. We capture this discrimination and unequal treatment in the toxicity of the environment measure in our model.

Early exposure to STEM has been shown to increase students' likelihood of pursuing a STEM degree [73]. One study suggests that female STEM students have a poorer understanding of what an ML/AI career looks like, which may contribute to lower rates of women entering ML/AI [24]. Therefore, we include a predictor in our model which measures exposure to the field.

"Data scientist" has been called "the sexiest job of the 21$^{st}$ century" [74]. A recent LinkedIn study found that "Machine Learning Engineer" was the fourth fastest growing job title in the USA [25]. This growth also brings high salaries; for example, Glassdoor reports an average base salary of approximately $155,000 USD for a Machine Learning Engineer in San Francisco, California [26]. The fast growth and high salaries attract many applicants to each job post, making the data science and ML/AI culture seem competitive [24]. In response to the demand for ML/AI jobs, universities increased the number of ML/AI courses offered [3]. The competitive, popular, and lucrative nature of ML/AI currently distinguish it from STEM in general, and we measure the influence of these factors in our work through three questions. First, we ask participants how important it is that they earn a high salary when compared to their skills and experience to gauge the relationship between a high salary and persistence in ML/AI. Second, prior research has examined differences in attitudes toward competitiveness [75]–[77], suggesting that men tend more, on average, toward harboring a "competing to win" mindset [76]. Researchers have theorized that competitiveness may shape interpersonal behaviors that influence school or work environments [77]. Given ML/AI's rapid growth as a field, we examine the possibility that those with a history of competitive participation (e.g., engaging in contexts with prizes or awards, such as athletics, judged performances, entrepreneurial competitions, etc.) may be comparably drawn toward working in ML/AI, and conversely, that those who have tended to avoid competitive environments may gravitate away from it. Third, we measure generally whether students were influenced to take the current ML/AI course due to the subject's overall popularity.

*Individual Characteristics*

Women have historically been underrepresented in STEM [78]. In fact, data from 2009 showed that gender was the most important variable in predicting an engineering major choice [79]. While women's participation in some STEM fields, such as in biology and chemistry, have increased significantly over the past 3 decades [78], [80], the share of women in engineering and computer science jobs remains stagnant at around 20% [80]. ML/AI faces similar rates of participation among women. According to the Global Gender Gap Report of 2021, women hold less than 25% of "AI Specialist", "Data Engineer", and "Big Data Developer" job titles [81]. According to Stanford's 2021 Artificial Intelligence Index Report, only 16% of tenure-track faculty at top universities globally, whose research area is AI, are women [3]. Studies on gender and persistence in STEM have found that, on average, women report lower levels of confidence in their abilities, career-fit, and sense of belonging in STEM, which correlates to lower levels of persistence in these fields [16], [43], [46]. For those who persist, however, recent work suggests that employers feel pressure to hire diverse candidates, as women engineering graduates are hired faster and for the same rate as men [82]. Our work continues the inquiry into how gendered patterns in confidence contribute to intentional persistence in the field of ML/AI specifically.

Racial and ethnic minorities, particularly Black, Hispanic, and Indigenous people, are also underrepresented in STEM. According to a 2021 PEW research report on diversity in STEM in the US, only 7% of professionals in computer jobs are Black and 8% are Hispanic [80]. Furthermore, the Brookfield Institute for Innovation and Entrepreneurship reports that Indigenous people account for less than 2% of technology workers in Canada and were 2 to 4 times less likely than non-Indigenous people to participate in technology [83]. Racial and ethnic minorities' persistence is influenced by stereotype threat [31], [84]. The engineering and STEM experience is not equal for all racial and ethnic minorities; for example, while Asian Americans are not considered underrepresented in STEM, they are impacted by unique stereotypes in engineering [85].

There are three additional student characteristics which we collect from our sample. First, student loan status, whether students took out loans to pay for school, is suggested to be an important factor to consider in our model. In general, studies have mixed results on the influence of student loan status on persistence [15], [28], [86], [87]. However, we can learn from past research which points to a relationship between student loan status and student risk orientation; this suggests that different careers carry different levels of financial risk. Such studies have found that engineering is seen as a low-risk, financially secure career path [30], [88], and thus we might expect to see a trend of increased engineering persistence with student loan status. Yet there do not exist studies of students' perceived risk of an ML/AI career path. Next, the persistence of first-generation students, students whose parents did not attend a four year university program, has also been studied [89]. They are another important group who have unique experiences in STEM [90][91]. Lastly, international student status, meaning a student who is not a citizen nor permanent resident of the country in which they attend university, is another characteristic that may be related to persistence in ML/AI. International students experience challenges due to cultural differences, social isolation, and language barriers [92]. International students find it more challenging to communicate in class [93] and experience more discrimination than domestic students [94] – both of which are factors thought to be associated with persistence. Thus, we include measures of student loan status, first generation student status, and international student status in our model.

**Methods**

We begin by overviewing our data collection methods and sample breakdown, followed by the measure of our dependent and independent variables, and conclude with a discussion of our analysis method.

*Data*

We conducted a self-selected survey study with undergraduate and graduate students enrolled in ML and/or AI courses in fall of 2021, at a major North American university. The study was approved by the university's Institutional Ethics Review Board. We targeted ML/AI courses that were introductory in nature and not narrowed to a specific application, in order to reach the largest audience. In total, we surveyed 10 courses within the departments of Computer Science and Engineering. Due to the hybrid nature of the courses during the semester surveyed, we administered surveys both in person (on paper) and digitally. We received a total of 191 responses, 77 from online respondents and 114 from in-person respondents. Response rates were much higher for those surveyed in-person, with a response rate of 87%, compared to 15% for those surveyed digitally.

Of the 191 surveys received, 165 were included in the sample after filtering for incomplete surveys (both dependent variables blank), those without a signed consent form, and those who indicated that they had taken the survey previously in another course. As seen in Table 1, 33% of our sample identified as women, which is similar to the 34.4% reported for undergraduate, and 27-34% reported for graduate engineering students at the university [95]. We also ask whether students identify as a visible minority, as per the government standardized demographic questions. This question lists 12 different visible minorities, along with "not a visible minority" and "other" options. Due to our sample size, only three of these categories had large enough sub-samples to include in the model. 79% of our sample identified as one or more visible minorities, which is slightly higher than the 69% of first years that identify as a visible minority, reported most recently by the university in 2017 [96]. In our sample, 42% identify as Chinese, 18% as South Asian, and 16% as not a visible minority, with a small number of students from each of the other minorities. Unfortunately, the university does not publish enrollment numbers for each visible minority, and thus we cannot analyze the representativeness of our sample for these variables. In 2019, the university reported that 32% of incoming undergraduate engineering students were international students, which is lower than the 44% of our sample that identified as international students. This statistic is not broken down by program, so it is possible that the computer science and engineering programs that we targeted attract a higher number of international students. In 2017, the university reported that 17% of incoming students were first-generation students [96], which is similar to the 13% found in our sample. The university does not publish figures on student loan status.

Graduate students make up 43% of our sample, followed by 33% in $3^{rd}$ year of undergrad, 22% in $4^{th}$ year, and 2% in $5^{th}$ year. 35% of our sample completed the survey online. Our participants represent a range of majors, from Mechanical and Industrial Engineering, Chemical Engineering, Computer Science, and non-Engineering majors. Due to the small number of Computer Science majors in our sample, we combined this category with Electrical and Computer Engineering, which is similar to the departmental grouping in many schools. We include the two largest major groups as controls in our model, 42% of our sample majored in Mechanical and Industrial Engineering and 25% in Electrical and Computer Engineering or Computer Science. Additionally, 57% of our sample indicated that they were enrolled in a ML/AI minor, specialization, or certificate, which can be taken in addition to any major.

*Dependent Variables*

Similar to Cech et al. [16] and Ren et al. [18], the dependent variable in this work is intentional persistence, meaning how likely it is that a student expects to remain in ML/AI in the future. We test two measures of intentional persistence: short-term and long-term. Short-term intentional persistence represents whether a student intends to take another ML/AI course in university; thus, only students who were not yet graduating responded to this question. Long-term intentional persistence represents whether a student intends to be in an ML/AI role in five years. As this study involved aggregating responses from both paper/in-person and online formats of the survey instrument, steps had to be taken to handle a few instances where survey questions were formatted differently between the two versions. Specifically, responses to the two dependent variable questions (short- and long-term intentional persistence) were measured on a five-point scale in the paper version and on a four-point scale in the online version. This required us to merge response types into a common format, while also checking that the conversion process did not substantively impact relationships between variables. We dichotomized the dependent variables into high-persistence and low-persistence based on the midpoint of the scale; responses less than or equal to two on the four- and five-point scale were labelled as low persistence, and responses of three or more on the four- and five-point scale were labelled as high persistence. A bivariate test for each dependent variables showed no significant difference between online and in-person responses, signaling that the original scale difference does not impact the dichotomized outcome (Appendix A).

*Independent Variables*

Building on the past work outlined in the previous section, we will be examining ten key independent variables, as well as individual-level characteristics, that we expect are associated with student persistence in ML/AI. Some variables have been shown to be associated with persistence in prior studies but are not hypothesized within the scope of this work. We measure expertise confidence ($\alpha = 0.767$) and career-fit confidence ($\alpha = 0.850$) as they are defined in [16]. Discrimination from teaching staff was shown to be negatively associated with persistence in past work [18]; here, we measure this discrimination as part of a scale representing the toxicity of the environment ($\alpha = 0.848$), containing discrimination in ML/AI courses, differences in treatment due to identity in ML/AI courses, and negative stereotype enforcement in ML/AI courses. All multi-item measures were created using an average of all included questions, and then were converted back to ordinal variables on the corresponding four- or five-point scale. This was done for ease of interpreting the odds ratios presented in the model results. To investigate whether the lucrative nature of ML/AI [25], [26], [74] may influence students' persistence, we asked respondents whether they agree (on a five-point Likert scale) that it is important to earn a high salary, relative to their skills and credentials. Also on a five-point scale, we asked respondents to rate how often they have recently participated in competitive events or activities to assess their competitive participation. To address the influence of the current popularity of ML/AI as a career path [25], we ask participants whether they agreed that the popularity of ML/AI influenced their choice to take the course, measured on a five-point scale. To measure the association of early exposure with persistence in ML/AI, we asked participants how long ago they learned of ML/AI as a career option. While our survey (fully reproduced in Appendix B) included other measures such as math and programming self-assessment and career identity, our sample size limited the number of variables we could include in the model. Math and programming are captured via the expertise confidence measure and are correlated with academic major, thus we did not include them in the model as individual variables.

The key independent variables that we examine in this work include non-technical self-assessment, social benefit interest, and our new measure social belonging confidence. Based on research that found gendered differences in how students rate themselves on non-technical skills [51], [52], as well as the observation that some business- or client-focused ML/AI roles have an emphasis on non-technical skills, we asked respondents to self-assess their communication, leadership, and teamwork skills compared to an average person their age. These three questions formed the non-technical self-assessment scale ($\alpha = 0.784$). Building on past work [18], we measured respondents' social benefit interest; how important it was that their work has a positive societal impact. This was measured using the altruistic work values scale from [32] ($\alpha = 0.809$). Lastly, we developed the social belonging confidence scale ($\alpha = 0.852$) to measure the degree to which a person feels that they will fit in with the social and cultural aspects of a profession and develop meaningful relationships with their peers. This scale consists of three questions relating to finding community, fitting into the professional culture, and relating to others, all measured on a four-point scale from not confident at all to very confident. The complete survey is included in Appendix B.

Lastly, we captured numerous demographic variables that have been shown to associate with student persistence or related outcomes, including gender, visible minority status [97], Indigenous identity [98], international student status, whether the student is a first-generation student, and whether they have student loans. It should be noted that we received no surveys from students who identified as Indigenous, and thus it is missing from the results section below. We also collected information about respondents' university program such as year of study, major, whether the student is enrolled in a ML/AI minor, specialization, or certificate, and whether they have taken a prior course in ML/AI.

*Analysis*

In preparing the survey data, we added control variables for survey modality (online vs. in-person). Appendix A shows that likely due to the self-selective nature of the online version of our survey, taking the survey online was more positively associated with the intention to persist in ML/AI both in the short-term and long-term than taking the survey in person. However, this difference was not significant (p = 0.701 in the short-term, p = 0.377 in the long-term). We also added a control for time of semester surveyed (mid-semester vs. late-semester). All surveys were conducted during the second half of the fall 2021 semester, so we used a 50/50 split of our data to denote mid-semester vs. end of the semester, as there were no notable events within this time to use as an appropriate split. Bivariate tests show that short-term persistence was approximately equivalent across time of semester (p = 0.887), but those who were surveyed at the end of the semester were marginally less likely to intend to remain in ML/AI in the long term (p = 0.088).

For the visible minority measure, we used the federal government definitions [97] to place any students who responded with a write-in value that belongs to one of the categories and created an additional category for those students who identify as two or more visible minorities. Due to the physical nature of the in-person surveys, for a number of questions we received some responses that did not fit within the provided scales. Some students wrote in values above, below, or in between scale values. If the value was above the maximum value on the scale or between two values, we rounded down, as a .5 rating suggests hesitancy and may inflate our results if we were to round up. In the cases where a response was below the minimum on the scale (e.g., a 0 on a 1-5 scale), we rounded up to the minimum value.

We used a multivariate logistic regression model and the statistical software Stata/BE version 17.0 to test our hypotheses. This model was run using the short-term intentional persistence dependent variable, and again using the long-term intentional persistence dependent variable. Independent variables and controls are identical across the two models, with the exception of South Asian visible minority which is removed in the short-term intentional persistence model due to perfect prediction. To build our models, the first iteration (step) of each model included controls, non-hypothesized independent variables, and gender. The second iteration added in predictors (hypothesized independent variables). The third model also added interaction effects between gender and independent variables that had significant bivariate tests, as shown in Table 3. We included only a binary variable for woman in our final models for interpretability, although hypotheses related to men are tested using bivariate tests with the variable "man" and included in Table 4. Thus, all independent variables that significantly differed between women and the rest of the sample are included as interaction effects in the third iteration of the models. The third iteration, Model 3, represents the most complete model of student persistence and is the model we use to test our hypotheses. For those independent variables that were not significant in the women vs. peers bivariate tests, but were significant in the men vs. peers bivariate tests, we included the *woman x independent variable* interaction effect in the model to test for significance. Only *woman x career-fit confidence* was significant and is included in the final model, while the rest are not. We also checked for differences in independent variable values by visible minority, international student status, and student loan status, which are in Appendix C, D, and E respectively.

*Results*

Table 1 shows a breakdown of our survey respondents, and Table 2 includes means and standard deviations for the independent variables measured on a scale. Table 3 includes means and standard deviations of variables comparing women to the rest of the sample, complete with bivariate tests. We had ten participants who did not identify as a woman or a man: four participants who identified as gender-fluid, non-binary, and/or Two-Spirit, and six who either preferred not to say or left the question blank. The significance tests shown in Table 3 compare those who identified as a woman to the rest of the sample, where the rest of the sample includes those who identify outside of the gender binary as well as those who preferred not to say. We also conducted these same significance tests comparing men to the rest of the group, and we marked with a dagger any variables in Table 3 where the men vs. peers test was significant but the women vs. peers test was not. Table 4 includes the bivariate test for men vs. peers used to address Hypothesis 3A.

From Table 3, we find evidence to support Hypothesis 1A, that women have higher levels of social benefit interest than their peers ($p < .01$). We do not find support for Hypothesis 2A, that women have higher levels of non-technical self-assessment than their peers. While the means reported in Table 3 align with the direction of our hypothesis, the sample size collected is not sufficient to show statistically significant support. We also find support for Hypothesis 3A, that men have higher levels of social belonging confidence than their peers ($p < .05$), as shown in Table 4.

Inconsistent with previous literature, we found no significant difference in short- or long-term intentional persistence by gender. We found that short-term intentional persistence is higher than long-term intentional persistence, suggesting that students are more certain that they will take another course in ML/AI than their intent to be in a ML/AI role in 5 years. We did not find gendered difference in expertise confidence levels, but we did find that men have significantly higher levels of career-fit confidence than their peers ($p < .05$). Overall, our participants indicated that the environment of their ML/AI courses was low in toxicity (note that this value is reverse scored in all tables).

Table 1: Summary Statistics

| Variable | Percent of Observations |
| --- | --- |
| Gender | |
|    Woman | 33.3 |
|    Man | 60.6 |
| Visible Minority | |
|    Chinese | 41.8 |
|    South Asian | 18.2 |
|    Not a visible minority | 15.8 |
| University Year | |
|    3rd year undergrad | 32.7 |
|    4th year undergrad | 21.8 |
|    5th year undergrad | 1.8 |
|    Graduate studies | 43.0 |
| Major | |
|    Mechanical and Industrial Engineering | 42.4 |
|    Electrical and Computer Engineering or Computer Science | 24.9 |
| ML Experience | |
|    Enrolled in ML/AI program | 57.6 |
|    Taken a ML/AI course prior | 50.3 |
| International student | |
|    Not a citizen or permanent resident of the country of their university | 44.2 |
| First generation student | |
|    Neither parent attended a four-year university program | 13.3 |
| Student loan status | |
|    Took out loans to pay for school | 32.7 |
| Survey modality | |
|    Took the survey online | 34.5 |
| Exposure to ML/AI | |
|    First learned about ML/AI as a career less than 1 year ago | 23.6 |
|    First learned about ML/AI as a career 1-3 years ago | 57.6 |
|    First learned about ML/AI as a career 4-5 years ago | 15.2 |
|    First learned about ML/AI as a career 6-9 years ago | 0.60 |
| Intentional Persistence in ML/AI | |
|    Intends to take another ML/AI course in university | 88.8 |
|    Intends to remain in a ML/AI role in five years | 74.4 |

In our sample, more women identified as Chinese than their peers, and more men identified as South Asian or not a visible minority than their peers. There is a higher percentage of men in graduate studies, men identifying as international students, and men who indicated that they have student loans. Approximately 50% of our sample have taken an ML/AI course prior to the course they were surveyed in. 58% of participants sampled indicated that they learned about ML/AI as a career path between one and three years ago, followed by 24% that selected 'less than a year ago,' suggesting that ML/AI is still a new career path that students learn about in university. Our participants indicated that earning a high salary relative to those with similar skills and experience was important, they reported moderate amounts of recent participation in competitive events or activities, and generally agreed that the popularity of ML/AI influenced their choice to take the course.

Table 2: Mean and standard deviations for independent variables measured on a scale.

| Variable | Mean | Standard deviation |
|---|---|---|
| Self-assessment | | |
|     Non-technical self-assessment | 3.747 | 0.775 |
| Confidence | | |
|     Career-fit confidence | 2.555 | 0.801 |
|     Expertise confidence | 2.921 | 0.767 |
|     Social belonging confidence | 2.630 | 0.795 |
| Other variables | | |
|     Social benefit interest | 4.389 | 0.724 |
|     Toxicity of environment[1] | 4.311 | 0.911 |
|     Importance of a high salary relative to skills and experience | 4.068 | 0.792 |
|     Competitive participation | 3.191 | 1.198 |
|     Influence of popularity of ML/AI on course choice | 3.660 | 0.992 |

[1] the values presented in the table for this variable are reverse scored; a high value for toxicity of environment in this table indicates a low level of toxicity.

Table 5 shows the results of the multivariate logistic regression, where columns labelled with A represent the prediction of short-term intentional persistence and those labelled with B represent the prediction of long-term intentional persistence. Models 1, 2, and 3 represent iterations of one model of student persistence, where Model 3 is the most complete model used to test our hypotheses. The first iteration of the model, Model 1, includes the gender coefficient (woman=1), individual controls, survey controls, and non-hypothesized independent variables such as career-fit confidence and expertise confidence. Consistent with the bivariate tests presented in Table 3, we do not find an association between women and persistence in either the short- or long-term. Though non-significant, the odds ratios throughout both short- and long-term persistence models are larger than one, suggesting that women may be more likely to persist in the field. We find that being a graduate student is a significant negative predictor of short-term intentional persistence, but not long-term intentional persistence.

Identifying as South Asian visible minority was removed from Model A due to perfect prediction, and we found no significant association of visible minority status with short- or long-term intentional persistence in any of the models. Identifying as an international student or having student loans did not significantly predict either short- or long-term intentional persistence; however, being a first-generation student was a borderline significant negative predictor of taking another ML/AI course.

Majoring in Electrical and Computer Engineering or Computer Science is a significant negative predictor of intending to take another ML/AI course; in fact, Model 3A shows that being in this major corresponds to a 97.8% decrease in odds that the student expects to take another ML/AI course. This major is also a borderline significant negative predictor of long-term intentional persistence. Being enrolled in an ML/AI program (specialization, minor, or certificate) was a borderline significant positive predictor of only short-term intentional persistence, just as having taken a prior course in ML/AI was a borderline significant positive predictor only of long-term intentional persistence. Being a graduate student has an odds ratio <1 throughout all models, but this ratio is very small and significant in the prediction of short-term intentional persistence.

Table 3: Bivariate tests comparing women with their peers

| Variable | Women (N=55) Mean (standard deviation) | Not Women (N=110) Mean (standard deviation) | $\chi^2$ sig. test |
|---|---|---|---|
| Visible Minority | | | |
|   Proportion Chinese | 0.618 | 0.318 | *** |
|   Proportion South Asian | 0.091 | 0.227 | * |
|   Proportion not a visible minority | 0.072 | 0.200 | * |
| University year | | | |
|   Proportion 3rd year | 0.400 | 0.291 | |
|   Proportion 4th year | 0.236 | 0.209 | |
|   Proportion 5th year | 0.018 | 0.018 | |
|   Proportion graduate studies | 0.345 | 0.473 | † |
| University program | | | |
|   Proportion Mechanical and Industrial Engineering major | 0.400 | 0.436 | |
|   Proportion Electrical and Computer Engineering or Computer Science major | 0.200 | 0.273 | |
|   Proportion ML/AI program | 0.618 | 0.555 | |
|   Proportion taken a prior course in ML/AI | 0.527 | 0.491 | |
| International student | | | |
|   Proportion international student | 0.364 | 0.482 | † |
| First generation student | | | |
|   Proportion first generation student | 0.109 | 0.145 | |
| Student loan status | | | |
|   Proportion student with loans | 0.200 | 0.391 | * |
| Exposure to ML/AI as a career option | | | |
|   Proportion < 1 year ago | 0.273 | 0.218 | |
|   Proportion 1-3 years ago | 0.582 | 0.573 | |
|   Proportion 4-5 years ago | 0.127 | 0.164 | |
|   Proportion 6-9 years ago | 0.000 | 0.009 | |
| Intentional persistence | | | |
|   Short-term intentional persistence | 0.939 (0.242) | 0.864 (0.344) | |
|   Long-term intentional persistence | 0.727 (0.449) | 0.752 (0.434) | |
| Self-assessment | | | |
|   Non-technical self-assessment | 3.811 | 3.716 | |

|  | (0.833) | (0.746) |  |
| --- | --- | --- | --- |
| Confidence |  |  |  |
|   Expertise Confidence | 2.891 | 2.936 |  |
|  | (0.832) | (0.736) |  |
|   Career-Fit Confidence | 2.400 | 2.633 | † |
|  | (0.760) | (0.813) |  |
|   Social Belonging Confidence | 2.491 | 2.697 | † |
|  | (0.823) | (0.776) |  |

Table 3 (continued): Bivariate tests comparing women with their peers

| Variable | Women (N=55) Mean (standard deviation) | Not Women (N=110) Mean (standard deviation) | $\chi^2$ sig. test |
| --- | --- | --- | --- |
| Other variables |  |  |  |
|   Competitive participation | 3.113 | 3.229 |  |
|  | (1.138) | (1.230) |  |
|   Influence of ML/AI popularity | 3.830 | 3.578 |  |
|  | (0.871) | (1.039) |  |

*$p < .05$; **$p < .01$; ***$p < .001$ † represents tests that were significant in the men vs. peers bivariate test, but not the women vs. peers bivariate test. [1]the values presented in the table for this variable are reverse scored. A high value for toxicity of environment in this table indicates a low level of toxicity. Note that Fisher's Exact Test was used for any Chi-Square test where at least one tabulated frequency was less than five [99].

In terms of survey controls, we found that responding to our survey online was a significantly positive predictor of short-term intentional persistence and a borderline significant positive predictor of long-term intentional persistence. Holding all else equal, those who took our survey online are approximately 126 times more likely to intend to take another ML/AI course (Model 3A) than those who took the survey in-person, which is likely due to the exaggerated self-selection effect from online respondents which is discussed more in the following sections. Being surveyed later in the semester was a significantly negative predictor of long-term intentional persistence.

Table 4: Bivariate tests comparing men and their peers

| Variable | Men (N=100) Mean (standard deviation) | Not Men (N=65) Mean (standard deviation) | $\chi^2$ sig. test |
| --- | --- | --- | --- |
| Confidence |  |  |  |
|   Social belonging confidence | 2.75 | 2.435 | * |
|  | (0.730) | (0.861) |  |

Consistent with previous work, we found that career-fit confidence and expertise confidence positively predict long-term intentional persistence and are borderline significant positive predictors of short-term intentional persistence. While the toxicity of the ML/AI course environment did not significantly predict intent to take another ML/AI course, it was a borderline significant negative predictor of long-term intentional persistence. This measure is reverse scored, which means that participants who rated their environment as less toxic, also indicated that they do not intend to remain in ML/AI in the long term. Early exposure to ML/AI as a career option and the importance of earning a high salary did not significantly predict either short- or long-term intentional persistence in any of the models. Interestingly, we found that participating in competitive activities was associated with decreased odds of intending to persist in ML/AI, significantly in the long-term, and borderline significantly in the short-term. The influence of the popularity of ML/AI on the choice to take a ML/AI course had opposite relationships with short- and long-term intentional persistence. In the short term, taking the course partly because of subject popularity was borderline significantly positively associated with the intent to take another course, but significantly negatively associated with the intent to remain in ML/AI long term.

The second iteration of the model, Model 2, adds our hypothesized variables: social benefit interest, non-technical self-assessment, and social belonging confidence. The third iteration of the model, Model 3, represents the attempt to enter seven interaction effects into the model, identified based on significant differences between genders shown in Table 3. However, only *student loan status x woman* could be entered into Model A, as the others resulted in perfect prediction, and only *student loan status x woman, Chinese x woman, and Not a visible minority x woman* were added to Model B. None of the interaction effects added to the model were significant. As Model 3 presents the most complete model of student persistence, this is the model we draw conclusions from in the following sections.

We find support for Hypothesis 1B that social benefit interest negatively predicts long-term intentional persistence ($p < .05$). The odds ratio shown in Models 3 suggests that a decrease of one ordinal value in social benefit interest results in a 71% increase in odds that the student will expect to work in ML/AI. We do not find support for Hypotheses 2B and 2C, as non-technical self-assessment does not significantly predict either short- or long-term intentional persistence in ML/AI. Although non-significant, in the short-term, the odds ratio is <1, which aligns with our hypothesis that high ratings of non-technical skills will negatively influence persistence. However, we see a borderline significant trend in the opposite direction for long-term intentional persistence; high ratings of non-technical skills are associated with the expectation to work in ML/AI in the future. Lastly, we also do not find support for Hypothesis 3B, that social belonging confidence will positively predict long-term intentional persistence. While non-significant, it appears that social belonging confidence has little to no association with intent to take another ML/AI course, but may actually be associated with decreased odds of expecting to work in ML/AI in the long-term.

**Methodological Limitations**

Because this study resides in the early stages of the research cycle, there are some limitations to the findings' generalizability. While students' persistence in traditional engineering pathways has been studied extensively, research on persistence in ML/AI is at a comparatively earlier stage. Thus, we prioritized establishing validity of the survey-based ML/AI career intentions data we collected. These decisions, as we next discuss, involved accepting trade-offs in sample size and sample recruitment for the present study, which allows us to advance a new survey instrument that can later be used in a more widescale survey deployment.

Table 5. Odds Ratios for the Multivariate Logistic Regression models predicting short-term intentional persistence (A) and long-term intentional persistence (B) in ML/AI. Models 2 and 3 are iterations on Model 1. Standard Errors are presented in parentheses.

| | Odds Ratios for Hypothesized Variables | | | | | |
| --- | --- | --- | --- | --- | --- | --- |
| | Model 1 | | Model 2 | | Model 3 | |
| Variable | A | B | A | B | A | B |
| Woman | 7.26 (9.032) | 1.307 (0.883) | 3.303 (4.744) | 1.760 (1.393) | 2.857 (4.206) | 2.331 (3.422) |
| Social belonging confidence | | | 1.031 (0.845) | 0.410 (0.225) | 1.074 (0.882) | 0.396 (0.228) |
| Social belonging confidence x woman | | | | | perfect prediction | perfect prediction |
| Non-technical self-assessment | | | 0.658 (0.577) | 2.713 (1.696) | 0.610 (0.559) | 3.201† (2.243) |
| Social benefit interest | | | 3.093 (2.528) | 0.350* (0.167) | 2.893 (2.416) | 0.293* (0.155) |
| Social benefit interest x woman | | | | | perfect prediction | perfect prediction |
| Career-fit confidence | 8.573* (8.099) | 5.15** (2.929) | 6.817† (7.397) | 9.541** (6.463) | 7.337† (8.117) | 12.768** (9.873) |
| Career-fit confidence x woman | | | | | perfect prediction | perfect prediction |
| Expertise confidence | 2.216 (1.864) | 6.947*** (3.632) | 1.861 (1.769) | 16.688*** (12.468) | 1.872 (1.78) | 21.338*** (17.605) |
| Toxicity of environment[1] | 1.694 (1.05) | 0.407* (0.174) | 1.833 (1.303) | 0.427† (0.197) | 2.052 (1.609) | 0.445† (0.218) |
| Early exposure | 0.468 (0.267) | 0.728 (0.241) | 0.425 (0.267) | 0.75 (0.271) | 0.402 (0.263) | 0.652 (0.256) |
| Competitive participation | 0.482† (0.212) | 0.689 (0.195) | 0.418† (0.221) | 0.564† (0.179) | 0.414† (0.218) | 0.506* (0.175) |

| | Odds Ratios for Hypothesized Variables | | | | | |
|---|---|---|---|---|---|---|
| | Model 1 | | Model 2 | | Model 3 | |
| Variable | A | B | A | B | A | B |
| Importance of high salary | 1.302 (0.71) | 1.208 (0.457) | 1.271 (0.753) | 1.289 (0.52) | 1.219 (0.74) | 1.257 (0.539) |
| Influence of ML/AI popularity | 2.762† (1.446) | 0.596 (0.204) | 2.894† (1.713) | 0.551† (0.195) | 2.899† (1.749) | 0.463* (0.18) |

Table 5 (continued). Odds Ratios for the Multivariate Logistic Regression models predicting short-term intentional persistence (A) and long-term intentional persistence (B) in ML/AI. Models 2 and 3 are iterations on Model 1. Standard Errors are presented in parentheses.

| | Odds Ratios for Hypothesized Variables | | | | | |
|---|---|---|---|---|---|---|
| | Model 1 | | Model 2 | | Model 3 | |
| Variable | A | B | A | B | A | B |
| Mechanical and Industrial Engineering major | 0.767 (0.87) | 0.262 (0.223) | 0.694 (0.834) | 0.300 (0.275) | 0.760 (0.922) | 0.319 (0.302) |
| Electrical and Computer Engineering or Computer Science Major | 0.028* (0.043) | 0.141* (0.132) | 0.021* (0.035) | 0.168† (0.16) | 0.022* (0.037) | 0.156† (0.151) |
| ML/AI program | 5.963† (6.139) | 1.241 (0.757) | 8.205† (9.41) | 1.641 (1.075) | 9.147† (11.196) | 1.948 (1.326) |
| Prior ML/AI course | 4.783 (4.903) | 2.96 (2.026) | 4.605 (5.077) | 3.986† (2.898) | 4.631 (5.086) | 3.895† (2.974) |
| Graduate student | 0.002** (0.004) | 0.429 (0.446) | 0.002** (0.005) | 0.317 (0.357) | 0.002** (0.005) | 0.316 (0.379) |
| International student | 2.200 (2.257) | 1.578 (1.123) | 1.220 (1.455) | 2.521 (1.927) | 1.312 (1.583) | 2.965 (2.426) |
| First-generation student | 0.166 (0.219) | 0.922 (0.785) | 0.08† (0.112) | 1.234 (1.146) | 0.07† (0.101) | 0.995 (0.918) |
| Student loan status | 0.132† (0.141) | 1.076 (0.813) | 0.14† (0.162) | 1.134 (0.933) | 0.323 (0.756) | 10.338 (17.757) |
| Student loan status =1 & woman=0 [2] | | | | | 0.357 (0.890) | 0.072 (0.137) |
| Chinese | 0.158† (0.171) | 0.749 (0.55) | 0.119 (0.158) | 0.856 (0.668) | 0.113 (0.154) | 0.538 (0.71) |
| Chinese=1 & woman=0 [2] | | | | | perfect prediction | 2.061 (3.273) |

| | Odds Ratios for Hypothesized Variables | | | | | |
| --- | --- | --- | --- | --- | --- | --- |
| | Model 1 | | Model 2 | | Model 3 | |
| Variable | A | B | A | B | A | B |
| South Asian | perfect prediction | 0.609 (0.611) | perfect prediction | 0.955 (0.997) | perfect prediction | 1.295 (1.455) |
| South Asian x woman | | | | | perfect prediction | perfect prediction |
| Not a visible minority | 0.884 (1.07) | 1.133 (1.046) | 0.683 (0.933) | 1.100 (1.164) | 0.724 (0.998) | 0.407 (0.899) |
| Not a visible minority=1 & woman=0 ² | | | | | perfect prediction | 3.331 (7.852) |

Table 5 (continued). Odds Ratios for the Multivariate Logistic Regression models predicting short-term intentional persistence (A) and long-term intentional persistence (B) in ML/AI. Models 2 and 3 are iterations on Model 1. Standard Errors are presented in parentheses.

| | Odds Ratios for Hypothesized Variables | | | | | |
| --- | --- | --- | --- | --- | --- | --- |
| | Model 1 | | Model 2 | | Model 3 | |
| Variable | A | B | A | B | A | B |
| Online | 175.364** (342.018) | 4.788 (5.246) | 113.439* (237.576) | 9.735† (11.899) | 125.691* (267.208) | 12.436† (16.743) |
| Late semester | 0.324 (0.361) | 0.146* (0.117) | 0.529 (0.646) | 0.061** (0.062) | 0.532 (0.648) | 0.046** (0.051) |
| Intercepts | 0.007 (0.03) | 1.492 (4.563) | 0.002 (0.01) | 0.722 (2.773) | 0.002 (0.010) | 0.790 (3.148) |
| Likelihood ratio chi-square statistic | 53.15*** | 89.98*** | 55.89*** | 91.12*** | 56.17*** | 93.58*** |
| Pseudo R-Squared | 0.505 | 0.468 | 0.531 | 0.508 | 0.533 | 0.522 |
| Total Observations | 147 | 159 | 147 | 159 | 147 | 159 |

†$p<.1$, *$p<.05$, **$p<.01$, ***$p<.001$. Total observations differ between models due to missing responses for some variables. Short-term models have fewer observations as some students were graduating and thus could not take additional courses. Models 1, 2, and 3 represent iterations of one model of student persistence, where Model 3 is the most complete model used to test hypotheses. Model 1 = controls + gender + other independent variables; Model 2 = Model 1 + predictors (social belonging confidence, non-technical self-assessment, social benefit interest); Model 3 = Model 2 + interaction effects between gender and independent variables. The first column in each model (A) predicts short-term intentional persistence, the likelihood of taking another ML/AI course in university. The second column in each model (B) predicts the long-term intentional persistence of being in an ML/AI role in 5 years. ¹the values presented in the table for this variable are reverse scored. ²Interaction effect relates to combination values shown as all other combinations were omitted for collinearity. The gender variable reported here represents women (coded as 1) compared with all other gender identities in the sample. We also tested the model with all gender identities encoded categorically and obtained the same statistical significance levels. The results are reported as shown for ease of interpretation of interaction effects with the gender variable.

Employing surveys to measure intentions data carries inherent validity risks [100], [101]. Individuals' intentions expressed on surveys may align poorly with subsequent real-life decisions if survey context, timing, and scope are not carefully selected. Kagan in 2017, for instance, discusses the risk of poor validity of surveys that ask respondents to envision being someplace or someone they are not, to assess threats or opportunities that are hypothetical, or to make decisions about situations that are far removed in time from respondents' present situations [101]. Asking respondents about their present situations and near-term decisions are more likely to yield valid responses. For these methodological reasons, we exclusively surveyed students presently enrolled in ML/AI courses who soon face decisions related to persistence in ML/AI. This approach constrained our possible sample in terms of access and the cooperation of their instructors to conduct the survey. Instructor-imposed constraints impacted survey time windows and how respondents could be recruited.

While this study's *in-situ* survey approach lends confidence toward the validity of the findings, it carries with it known adverse impacts on sample size and response rate, which, in turn, limit our claims of generalizability to the broader population of all prospective candidates for careers in ML/AI. Within the imposed constraints (e.g., extent of class time or online platform presence granted, and the extent of requests or encouragement to complete the survey), this study achieved a 23.6 % response rate and sample size of 165 respondents. The response rate (in the absence of random assignment) suggests that participants self-selected into the survey. A self-selected sample in a study on career intentions could disproportionally over- or under-represent attitudes about working in a particular field; for instance, those with stronger opinions could be more likely to respond. Therefore, while the present study allows for meaningful comparative analyses based on within-sample respondent characteristics, it does not allow us to quantify the extent to which the salience of relationships between key independent and dependent variables maps to the broader population. Further, our survey deployment strategy and limited resources restricted our ability to recruit a larger sample for this study. The present sample size was sufficient to enable detection of statistically significant relationships in our statistical models but was insufficient to provide a completely cross-sectionally representative sample with respect to participant demographics (i.e., based on the population demographics of Engineering and Computer Science students at the university [95], [96]). This early-stage study must therefore be interpreted with caution: our findings reveal key relationships among variables pertinent to the population of interest, but do not allow us to verify population-level claims about trends underlying widescale persistence or attrition behaviors observed among ML/AI career candidates. Follow-on research with a demographically representative and, to the extent possible, non-self-selected sample are required to assess population-level generalizability. We hope that the survey measures developed in this present study can be leveraged in such future work.

**Discussion and Future Work**

Our survey of undergraduate and graduate level students taking ML/AI courses revealed several key variables that associate with students' short- and long-term intentional persistence in the field. Short-term intentional persistence, whether students expect to take another ML/AI course in university, is mostly associated with characteristics of their enrollment, such as level of study and major. We found that long-term intentional persistence in ML/AI is positively predicted by the professional role confidence measures from [16], along with some evidence that high confidence in non-technical skills and having taken a prior course in ML/AI are positively associated with long-term intentional persistence. High values of social benefit interest, frequently participating in competitive activities, course choice being influenced by the popularity of ML/AI, and being surveyed late in the semester were negatively associated with expecting to work in ML/AI. We discuss the implications of each of these key variables in the following paragraphs.

Perhaps unsurprisingly, short-term intentional persistence is mostly associated with academic enrollment variables and has fewer significant predictors overall. Enrollment in a specific ML/AI program, which we would expect to be a positive predictor of short-term intentional persistence, is only borderline significant. This may be caused by student curriculum planning; there may be varying requirements for ML/AI "add-on" programs, such as certificates or minors, which can be taken within any major. If only few courses are required for the certification, students taking their last required course may not want to take any additional unrequired courses or may have already taken all offered ML/AI courses. Being a graduate-level student is also a negative predictor of taking another course in ML/AI, which may suggest that graduate students enroll in ML/AI courses for breadth rather than their specialization, or that there may be a smaller number of courses graduate students can take in their degrees.

There are many borderline significant variables in the short-term model that should be further examined in future work. We find that career-fit confidence is a borderline significant positive predictor of short-term intentional persistence. As it has been shown to predict long-term intentional persistence in ML/AI [18], we expect that it would also be significant in the short-term with a larger sample size. Notably, the influence of ML/AI popularity on course choice is an interesting key variable that may shed light on how students choose ML/AI courses. Taking an ML/AI course due to the subject's overall popularity was a positive, although only borderline significant, predictor of short-term intentional persistence, suggesting that the popularity of ML/AI would continue to influence students to take more courses after they had taken the first one. On the other hand, it is a significant and negative predictor of long-term intentional persistence. These findings together suggest that the current rapid growth of demand for AI professionals [25] may influence course choice, but not overall career choice.

While the short-term intentional persistence model provides some insight into student course choice, the long-term intentional persistence model provides a more detailed understanding of how students may perceive ML/AI careers. One of the most striking findings is the negative association of social benefit interest with long-term intentional persistence, which aligns with the previous work in ML/AI [18]. We expanded this measure from a single question to a multi-item pre-validated measure from Lyons et al. [32], which shows that this effect is robust to the way we measured social benefit interest. [32]. Our improved measure also shows that women are more interested in work with social benefits than their peers, an effect which was previously non-significant using a single question measure [18]. This suggests a potential driver of the gender gap in ML/AI; women are interested in careers that positively benefit society and may not view ML/AI as one. For example, studies have shown that women are more worried about personal data collection and reduced human connection as impacts of AI use [22], and are less enthusiastic about AI use overall [22], [23], [42]. While ML/AI may be negatively portrayed in the media, ML/AI educators can focus on the opportunities it has to improve society, as outlined in [39]: enabling human self-realization, enhancing human agency, increasing societal capabilities, and cultivating social cohesion. Recent research suggests that making students aware of the meaningful nature of work in ML/AI and computer science will help retain women [21]. Educators can also highlight socially beneficial applications of ML/AI, such as cancer detection algorithms [102], but should not paint an overly positive image. The inherent risks of ML/AI must still be covered in these courses, and these lessons must go beyond simply teaching technical "fixes" [103].

Consistent with previous work in engineering [16] and ML/AI [18], we found that even given our limited sample size, professional role confidence, consisting of career-fit confidence and expertise confidence measures, was a strong positive predictor of long-term intentional persistence. This further suggests that being confident that a role will fit into your needs and interest, as well as being confident that you have the skills and competencies needed for the role, are important in persistence. This finding provides additional evidence to suggest that engineering persistence research can also apply to the new and developing field of ML/AI and justification for borrowing and building from this vast body of literature. In addition to the professional role confidence measures, we developed the social belonging confidence measure to capture the degree to which a person feels that they will fit in with the social and cultural aspects of a profession and develop meaningful relationships with their peers. While we hypothesized that this would positively associate with long-term persistence in ML/AI, the resulting odds ratios are in the opposite direction, wherein high levels of social belonging confidence may be associated with not expecting to work in ML/AI. This measure was not a significant predictor in our model and should be addressed further in future work. While we demonstrated internal consistency in this work, the next iteration of this study will seek to measure the construct validity. Specifically, we are currently supplementing this data by interviewing ML/AI students about their experience in classes, research, or internships, and how this impacts their sense of belonging within the community of ML/AI. Through this work we hope to gain insight on how social belonging confidence influences students' persistence in the field and how intersectional gender-racial identities shape students' experiences.

The toxicity of environment measure should also be externally validated in future work. While we expect that students who have experienced discrimination, unequal treatment, and the enforcement of negative stereotypes in their ML/AI courses would be less likely to expect to remain in the field long-term, we actually see a borderline significant trend in the opposite direction. While this could be explained by students not believing that their negative experiences in ML/AI courses will continue in a professional ML/AI role, this likely suggests that the measure construction should be revisited. This is further supported by the fact that discrimination from teaching staff, which is a component of our measure, was a significant negative predictor in past work [18] and has been shown to impact women's job success in STEM [70][64].

Our study also provides insight on the perception of the types of skills that are thought to be required and valued in ML/AI roles. While the expertise confidence measure asks participants whether they believe they have the skills to succeed in ML/AI, it is not clear what these skills are. While math skills are often included in STEM [46] and engineering [16] persistence studies, interpersonal skills are often not, despite their requirement in the workplace [19]. We measured non-technical self-assessment, and based on past work which suggested that the difference in technical and non-technical skills is a driver of the gender gap in STEM fields [51], [52], we expected that high levels of non-technical self-assessment would be negatively associated with intentional persistence in ML/AI. While this directionality was reflected in short-term intentional persistence, we see a borderline significant opposite relationship for long term persistence; being confident in your interpersonal skills is positively associated with expecting to work in ML/AI. This suggests that students are aware that these jobs require more than just programming and math skills, and educators should continue to emphasize this. While non-significant in this sample, the directionality of the bivariate test shows that women may have higher levels of confidence in non-technical self-assessment, and thus emphasizing the importance of these skills in ML/AI roles may help to close this gap. Educators should also design courses that help develop these important interpersonal skills through team projects and presentations. We found that specifically those who identified as Chinese had lower levels of non-technical self-assessment than their peers, so building these skills in classes may also help to support racialized students as well.

While we did not explicitly include a measure of technical skills in our model, we can reason about these skills based upon departmental affiliation. We expect, for instance, that those enrolled in a programming-heavy major likely have comparably strong programming skills. Yet, we found being enrolled in Electrical and Computer Engineering or Computer Science to be a strong negative predictor of intending to take another ML/AI course, and a slightly less significantly negative predictor of intending to persist in ML/AI in the long term. These findings suggests that the relationship between computer programming skills, departmental affiliation, and persistence in ML/AI is more complex than can be measured with a binary prediction variable and should be examined with a more robust skill measure in future work. Findings also highlight the importance in future studies of capturing where students intend to work if they do not intend to remain in ML/AI; it is plausible that Electrical and Computer Engineering students indicating plans to work outside of ML/AI intend to remain in a related technical position, such as in software development. In future work, we will ask participants specifically what role they expect to have in the future.

Interestingly, the other sizable major included in our sample, Mechanical and Industrial Engineering, was also a negative predictor of intentional persistence in ML/AI in both the short- and long-term, suggesting a need for a closer examination into which majors tend toward persisting in ML/AI. While the groups were too small to analyze in this model, we found that students who intended to persist in both the short- and long-term came from a multitude of ML/AI related majors such as Machine Intelligence or Data Analytics, as well as other engineering disciplines such as Chemical Engineering or Financial Engineering. The former group's intentions to work in the field in the future are understandable, as they have dedicated their degree to its study. The latter group, however, likely had to go out of their way to enroll in a ML/AI course, since such is not typically part of their department's curriculum. This latter group's departments also may not widely advertise ML/AI courses as electives compared to departments such as Electrical and Computer Engineering, Computer Science, or Mechanical and Industrial Engineering. Thus, these students likely held prior interest in the topic and may have chosen to take the course to help prepare for a future in the field. While not included in this model, other career identity information collected from participants (Appendix F) can help with interpretation of these findings. Of the options Engineer, Computer Scientist, Data Scientist/Data Analyst/Business Intelligence Professional, or Other, students were asked to select which role within ML/AI they would most closely identify with if they remained in the field. We can see that those in Electrical and Computer Engineering or Computer Science and Mechanical and Industrial Engineering identified more as an Engineer, while other majors identified in the Data Scientist/Data Analyst/Business Intelligence Professional category. This suggests that there are nuanced career identities within ML/AI which may relate to different factors that influence persistence and should be further examined in future work.

We posited that the current high growth, high salaries, and popularity of ML/AI [25], [26] are characteristics that distinguish the field from engineering and STEM in general. Specifically, we asked participants whether they were influenced to take the course due to the popularity of ML/AI as a subject, how important a relatively high salary was to them, and whether they frequently engage in competitive activities. We found, contrary to what we might expect, that frequent engagement in competitive activities is a significant or borderline significant negative predictor for both short- and long-term intentional persistence in ML/AI. One potential explanation could be that the high demand and fast growth in the field suggest that work opportunities are plentiful, therefore attracting individuals less interested in needing to compete for positions. While our measure of competitive participation as a proxy for competitiveness has limitations, competitiveness measured using a different survey question was shown to be a strongly significant negative predictor of long-term intentional persistence in ML/AI in past work as well [18]. The importance of a high salary, meanwhile, was non-significant across all models, but exhibited the expected effect direction (i.e., that importance of earning a high salary is associated with intentional persistence in ML/AI). Lastly, we examined the influence of ML/AI popularity on course choice. In general, our participants indicated they were somewhat influenced to take the course due to the popularity of the subject, and while this positively predicted that they expected to take another ML/AI course, it was a negative predictor of long-term intentional persistence. These findings suggest that taking an ML/AI course due to external influence, as opposed to internal interest and desire, may be negatively associated with intentional persistence. When measuring persistence and retention in this field in the future, researchers should bear this in mind: studies should account for the possibility that a subset of students are taking an ML/AI course in part due to its popularity.

The current analysis failed to uncover many relationships between individual-level characteristics and intentional persistence. While gender was a significant predictor of intentional persistence in past work [16], [18], it was non-significant across all of our models. However, its direction suggests that women may be more likely to intend to persist in ML/AI, which is consistent with past work in ML/AI [18], but inconsistent with past work in engineering [16]. We also did not find any effect of visible minority status on intentional persistence, which may be due to the limited sample size of each group. International student status, first-generation student status, and student loan status also were not significant predictors of either short- or long-term intentional persistence.

However, bivariate tests of these individual-level characteristics revealed intriguing patterns that can motivate future work. While testing hypotheses encompassing demographic attributes other than gender was outside of the scope of this paper, we include bivariate tests for visible minority, international student status, and student loan status in the appendices (Appendices C, D, E). Among these findings, we observed that those who identified as South Asian were more likely to intend to take another ML/AI course and generally had higher levels of expertise confidence. In contrast, those who identified as Chinese had, on average, lower levels of expertise confidence and non-technical self-assessment, and less frequently participated in competitive activities when compared to their peers. Those who identified as not a visible minority rated themselves higher, on average, on the non-technical self-assessment, rated their environment as significantly less toxic, and indicated that they more frequently participated in competitive activities compared to their peers. As such factors are believed to relate to persistence in ML/AI, these findings suggest that it is worthwhile to collect a larger and more diverse dataset to investigate how persistence may vary across demographic groups.

Our findings also that suggest that the role of international student status warrants further investigation toward strengthening our understanding of persistence. In this present study, international students were more likely to intend to be in a ML/AI role in five years, but they also reported learning about ML/AI as a career choice more recently than their peers. Further, they reported higher levels of social benefit interest than their peers, suggesting that the relationship between social benefit interest and persistence in ML/AI may be more complex in this group. As we further this work with a larger sample size, we will continue to investigate the salience of international student status.

Finally, in interpreting the present study, it is important to consider possible implications underlying the significance of certain survey control variables. Taking our survey online was a strong positive predictor of short-term intentional persistence, and a borderline significant positive predictor of long-term intentional persistence. This phenomenon likely relates to the self-selective nature of online survey participation. In the case of online classes, professors sent an announcement with the survey link, which elicited a comparatively low response rate compared to that from in-person classes. It is plausible that the self-selected subset who chose to respond online have comparably stronger interests or opinions related to ML/AI courses and careers than their peers, and therefore represent students who are comparably more likely to want to take another ML/AI course and remain in the field. As universities return to in-person courses, we will continue this work by surveying exclusively in-person to avoid this effect in the future. Additionally, we found that surveying students later in the semester was a significantly negative predictor of long-term intentional persistence. While this context-dependent effect is empirically nonideal, it shed light on an important finding. Coupled with our data that shows many students are influenced to take ML/AI courses due to the general popularity of the topic, it seems that students may not understand the type of work required in this field, and after they experience projects and assignments in the course, they realize it may not be the best fit. However, this could also be caused by a loss in confidence in their skills after receiving a poor grade or completing a tough assignment. Either way, the specific mechanisms underlying this late semester effect should be addressed in future work.

Based on the preliminary findings presented in this work, we plan to revise our survey to better measure where students plan to go if they do not plan to persist in ML/AI. We will distribute this survey broadly at multiple universities and collect a larger sample to both validate the present findings and to formally test additional variables. Measures that were included in the model but not formally hypothesized in this work can be the focus of future iterations of this study. Further, intentional persistence is only one piece of the broader persistence measure; to better understand persistence, we also need to measure how many students actually persist in ML/AI in the future (behavioural persistence). Participants who responded to this survey had the option to opt-in to be contacted for a follow-up survey in five years, and we will continue to request this permission from future participants as we iterate on this work.

The findings presented in this work have implications for ML/AI educators who hope to encourage persistence and help to increase diversity in the field. Educators should emphasize the positive benefits that ML/AI can have on society and its ethical implications to ensure that students can apply ML/AI in ways that align with their values. Educators should also encourage the development of interpersonal skills in these courses, through team projects, presentations, and leadership opportunities. For example, they can host guest speakers who work in ML/AI in industry to describe how they use interpersonal skills in their role. While we did not find social belonging confidence to be a significant predictor of persistence in this work, past work found that interventions aimed at increasing the feeling of social belonging in engineering may be a means of increasing diversity [104]. In addition to making changes to the way that we teach ML/AI, we must also pay attention to *who* is teaching it. Previous research on closing the gender gap in engineering found that structural changes in hiring processes and admission procedures compound over time to produce positive change [105], and departments that award more degrees to students identifying as underrepresented minorities are more likely to hire underrepresented minority women faculty [106]. Lastly, educators should be aware that many students are currently taking ML/AI courses to gain skills in a popular area, but they may not have the intention of working in the field long-term.

**Conclusion**

The new field of ML/AI lags far behind other STEM fields in terms of diversity. In this work, we present the advancement of a model that examines factors associated with intentional persistence of students in ML/AI in order to identify areas for improvement that can increase diversity in the field. We conducted a survey of undergraduate and graduate students enrolled in ML/AI courses at a major North American university in fall 2021. Our findings suggest that the intention to take another ML/AI course is associated with academic enrollment factors such as major and level of study. We found that measures of professional role confidence originally developed to study persistence in engineering are also important predictors of intent to remain in ML/AI. Unique to our study, we show that wanting one's work to have a positive social benefit is a negative predictor of long-term intentional persistence in ML/AI, and that women generally care more about this. Additionally, we found that having high confidence in non-technical interpersonal skills to be a positive predictor of long-term intentional persistence in the field. We provide recommendations to educators to emphasize the positive benefits that ML/AI can have on society and to encourage the development of interpersonal skills. Improving diversity in the field is a critical step in ensuring that the decisions made by algorithms represent the diverse set of users they will impact.

**Acknowledgement**

We would like to acknowledge the generous course instructors who invited us into their classrooms, and all the students who completed our survey. We would also like to acknowledge Prachi Sukhnani for her help developing the survey.

Reasonable inquiries for use of this dataset for research which adhere to our Research Ethics Board agreement will be accommodated by the authors; please contact the corresponding author for such requests.

**Appendices**

Appendix A: Associations between survey modality on outcome variable

| Modality | Short-term intentional persistence | | | Long-term intentional persistence | | |
|---|---|---|---|---|---|---|
| | Mean | SD | Test Statistic | Mean | SD | $\chi^2$ Test Statistic |
| Online | 0.901 | 0.300 | 0.701 | 0.786 | 0.414 | 0.377 |
| In-person | 0.881 | 0.325 | | 0.722 | 0.450 | |

*p < .05; **p < .01; ***p < .001.

Appendix B: Survey Questions

| *Survey Questions* | *Variable measured* |
|---|---|
| 1. In what major are you enrolled? *Open-ended* | N/A |
| 2. Are you enrolled in a ML/AI specialization/major/minor/certificate?  ☐ Yes  ☐ No  If yes, which one? *Open-ended* | N/A |
| 3. Have you previously taken another ML/AI course prior to this one?  ☐ Yes  ☐ No  ☐ Unsure | N/A |
| 4. How long ago did you learn about Machine Learning/Artificial Intelligence as a career option? | *Exposure to ML/AI* |

- ☐ Less than 1 year ago
- ☐ 1 – 3 years ago
- ☐ 4 – 5 years ago
- ☐ 6 – 9 years ago
- ☐ 10+ years ago

*Questions 5-6 were measured on a 5-point Likert scale from "Very Unlikely" to "Very Likely"*

5. What is the likelihood that you will take another ML/AI course in university? — *Short-term Persistence*
6. What is the likelihood that you will be in an ML/AI role (academia or industry) in 5 years? — *Long-term Persistence*

*Questions 7-16 were measured on a 4-point Likert scale from "Not confident at all" to "Very Confident"*

7. ML/AI is the right profession for me. — *Career-fit Confidence*
8. I can select the right role in ML/AI for me.
9. I can find a satisfying job in ML/AI.
10. I am committed to ML/AI, compared to my ML/AI classmates

11. I will develop useful skills through working with ML/AI. — *Expertise Confidence*
12. I will advance to the next level of my career in ML/AI.
13. I have the ability to be successful in my career in ML/AI.

14. I will find community in the field of ML/AI. — *Social belonging Confidence*
15. I will fit in with the professional culture in the field of ML/AI.
16. I will be able to relate to others in the ML/AI professional community.

*Questions 17-20 were measured on a 5-point rating scale from "Lowest 10%" to "Highest 10%"*

17. Rate your math ability compared to an average person your age — *Self-assessment Technical*
18. Rate your programming ability compared to an average person your age.

19. Rate your communication skills (e.g. writing and presenting) compared to an average person your age — *Self-assessment Non-technical*
20. Rate your teamwork skills compared to an average person your age.
21. Rate your leadership abilities (eg. planning, delegating, and coordinating) compared to an average person your age.

*Questions 22-30 were measured on a 5-point Likert scale from "Strongly Disagree" to "Strongly Agree"*

22. It is important to me to do work that makes a helpful contribution to society; makes a difference. — *Social Benefit Interest*
23. It is important to me to do work that is consistent with my moral values.

24. It is important to me to work in an environment where workplace policies are administered with fairness and impartiality.

25. It is important to me that I earn a high salary (I.e., high relative to typical salaries for those with my skills and credentials) in my career. *Earning Potential*

26. Now or in the recent past, I choose to participate in competitive events or activities (for instance: competitive athletics, judged performances, contests for funding or awards, entrepreneurial competitions, etc.)** *Competitive Participation*

27. I was influenced to take this course because of the popularity of ML/AI as a topic of study. *Interest in ML/AI*

28. I have experienced discrimination in some form in my ML/AI courses, in ways such as, but not limited to: *Toxicity of Environment*
    - Direct (unequal treatment based on race, colour, sex, etc.)
    - Indirect or Systemic (driven by discriminatory policies or practices)
    - Harassment (unwelcome comments or actions)
    
    Which had the impact of excluding me, denying me benefits, or imposing a burden on me

29. I have noticed differences in the way I am spoken to in my ML/AI courses compared to my peers of a different identity.

30. I can identify instances in my ML/AI courses where negative stereotypes regarding my identity, academic standards for my identity, and/or expectation of ability of my identity, were reinforced.

31. During and after my undergrad, if I remain in ML/AI (either in industry or academia), I will identify as: *Career Identity*
    - ☐ Engineer
    - ☐ Computer Scientist
    - ☐ None of the above
    - ☐ Data Scientist/Analyst/Business Intelligence
    - ☐ Other (please specify)

32. In one sentence, why do you categorize yourself the way that you did above? *Open-ended*

33. What year of study are you in? *Year of Study*
    - ☐ 1st year undergraduate
    - ☐ 2nd year undergraduate
    - ☐ 3rd year undergraduate
    - ☐ 4th year undergraduate
    - ☐ 5th year undergraduate
    - ☐ Master of Engineering student
    - ☐ Master of Applied Science/ Master of Science student
    - ☐ Doctor of Philosophy student

34. Do you identify as a visible minority in Canada?  *Race & Ethnicity*
    ☐ South Asian      ☐ Chinese
    ☐ Southeast Asian  ☐ Black
    ☐ Filipino         ☐ Latin American
    ☐ Arab             ☐ West Asian
    ☐ Korean.          ☐ Japanese
    ☐ Other: _______________________________
    ☐ Not a Visible Minority
35. Do you identify as an Aboriginal person, that is, First Nations (North American Indian), Métis, or Inuk (Inuit)?
    ☐ Yes  ☐ No  ☐ Prefer not to say
36. What best describes your gender?  *Gender Identity*
    ☐ Woman  ☐ Man  ☐ Genderfluid/Non-binary/Two-Spirit
    ☐ Prefer not to say
37. Are you an international student (i.e., not a permanent resident nor citizen of the nation of your college/university)  *International Student Status*
    ☐ Yes  ☐ No  ☐ Prefer not to say
38. Are you a first-generation student (neither of your parents have obtained a four-year college/university degree)  *First Generation Student Status*
    ☐ Yes  ☐ No  ☐ Prefer not to say
39. Did you or your family take out loans to pay for your college/university?  *Student Loan Status*
    ☐ Yes  ☐ No  ☐ Prefer not to say

---

*Variables are for informational purposes only and were not printed on the surveys
**This question was measured on a 5-point Likert scale from "Never" to "Very Often".

Appendix C: Bivariate Tests by race

Table 6: Bivariate Tests for South Asian visible minority

| Variable | South Asian (N= 30) Mean (standard deviation) | Not South Asian (N = 135) Mean (standard deviation) | $\chi^2$ sig. test |
|---|---|---|---|
| Proportion women | 0.167 | 0.370 | * |
| Proportion man | 0.800 | 0.563 | * |
| Proportion 3rd Year | 0.267 | 0.341 | |
| Proportion 4th Year | 0.267 | 0.207 | |
| Proportion 5th Year | 0.000 | 0.022 | |
| Proportion graduate studies | 0.467 | 0.422 | |
| Proportion Mechanical and Industrial Engineering major | 0.400 | 0.430 | |
| Proportion Electrical and Computer Engineering or Computer Science major | 0.233 | 0.252 | |
| Proportion ML specialist/minor/certificate | 0.667 | 0.556 | |
| Proportion taken a prior course in ML/AI | 0.633 | 0.474 | |
| Proportion international student | 0.300 | 0.474 | |
| Proportion first generation student | 0.133 | 0.133 | |
| Proportion student with loans | 0.633 | 0.259 | *** |
| Proportion < 1 year ago | 0.200 | 0.244 | |
| Proportion 1-3 years ago | 0.633 | 0.563 | |
| Proportion 4-5 years ago | 0.167 | 0.148 | |
| Proportion 6-9 years ago | 0.000 | 0.007 | |
| Short-term intentional persistence | 1.000 (0.000) | 0.863 (0.345) | * |
| Long-term intentional persistence | 0.833 (0.379) | 0.724 (0.449) | |
| Non-technical self-assessment | 3.800 (0.887) | 3.734 (0.750) | |
| Expertise Confidence | 3.400 (0.621) | 2.813 (0.758) | ** |
| Career-Fit Confidence | 2.833 (0.791) | 2.493 (0.792) | |
| Social Belonging Confidence | 2.967 (0.765) | 2.553 (0.785) | |

| Variable | South Asian (N= 30) Mean (standard deviation) | Not South Asian (N = 135) Mean (standard deviation) | $\chi^2$ sig. test |
|---|---|---|---|
| Social Benefit Interest | 4.433 (0.626) | 4.379 (0.747) | |
| Toxicity of environment | 4.467 (0.681) | 4.276 (0.953) | |
| Importance of high salary | 4.138 (0.743) | 4.053 (0.804) | |
| Competitive participation | 3.200 (1.157) | 3.189 (1.211) | |
| Influence of ML/AI popularity | 3.800 (0.847 | 3.629 (1.022) | |

*p < .05; **p < .01; ***p < .001 (two-tailed test).
Note that Fisher's Exact Test was used for any Chi-Square test where at least one tabulated frequency was less than 5 [99]

Table 7: Bivariate test for Chinese visible minority

| Variable | Chinese (N=69) Mean (standard deviation) | Not Chinese (N=96) Mean (standard deviation) | $\chi^2$ sig. test |
|---|---|---|---|
| Proportion women | 0.493 | 0.219 | *** |
| Proportion man | 0.478 | 0.698 | ** |
| Proportion 3rd Year | 0.304 | 0.344 | |
| Proportion 4th Year | 0.217 | 0.219 | |
| Proportion 5th Year | 0.014 | 0.021 | |
| Proportion graduate studies | 0.464 | 0.406 | |
| Proportion Mechanical and Industrial Engineering major | 0.333 | 0.490 | * |
| Proportion Electrical and Computer Engineering or Computer Science major | 0.290 | 0.219 | |
| Proportion ML specialist/minor/certificate | 0.609 | 0.552 | |
| Proportion taken a prior course in ML/AI | 0.493 | 0.510 | |
| Proportion international student | 0.493 | 0.406 | |
| Proportion first generation student | 0.130 | 0.135 | |
| Proportion student with loans | 0.232 | 0.396 | * |
| Proportion < 1 year ago | 0.232 | 0.240 | |

| Variable | Chinese (N=69) Mean (standard deviation) | Not Chinese (N=96) Mean (standard deviation) | $\chi^2$ sig. test |
|---|---|---|---|
| Proportion 1-3 years ago | 0.609 | 0.552 | |
| Proportion 4-5 years ago | 0.116 | 0.177 | |
| Proportion 6-9 years ago | 0.000 | 0.010 | |
| Short-term intentional persistence | 0.857 (0.353) | 0.910 (0.288) | |
| Long-term intentional persistence | 0.725 (0.450) | 0.758 (0.431) | |
| Non-technical self-assessment | 3.552 (0.744) | 3.884 (0.770) | * |
| Expertise Confidence | 2.739 (0.670) | 3.053 (0.790) | * |
| Career-Fit Confidence | 2.391 (0.712) | 2.674 (0.844) | |
| Social Belonging Confidence | 2.522 (0.766) | 2.705 (0.810) | |
| Social Benefit Interest | 4.358 (0.732) | 4.411 (0.722) | |
| Toxicity of environment | 4.203 (0.933) | 4.389 (0.891) | |
| Importance of high salary | 4.104 (0.761) | 4.043 (0.815) | |
| Competitive participation | 2.896 (1.233) | 3.400 (1.134) | * |
| Influence of ML/AI popularity | 3.866 (0.815) | 3.516 (1.080) | |

*p < .05; **p < .01; ***p < .001 (two-tailed test). Note that Fisher's Exact Test was used for any Chi-Square test where at least one tabulated frequency was less than 5 [99]

Table 8: Bivariate tests for "not a visible minority".

| Variable | Not a visible minority (N=26) Mean (standard deviation) | Did not select "Not a visible minority" (N=139) Mean (standard deviation) | $\chi^2$ sig. test |
|---|---|---|---|
| Proportion women | 0.154 | 0.367 | * |
| Proportion man | 0.769 | 0.576 | |
| Proportion 3rd Year | 0.423 | 0.309 | |

| Variable | Not a visible minority (N=26) Mean (standard deviation) | Did not select "Not a visible minority" (N=139) Mean (standard deviation) | $\chi^2$ sig. test |
|---|---|---|---|
| Proportion 4th Year | 0.192 | 0.223 | |
| Proportion 5th Year | 0.077 | 0.007 | |
| Proportion graduate studies | 0.308 | 0.453 | |
| Proportion Mechanical and Industrial Engineering major | 0.615 | 0.388 | * |
| Proportion Electrical and Computer Engineering or Computer Science major | 0.192 | 0.259 | |
| Proportion ML specialist/minor/certificate | 0.654 | 0.561 | |
| Proportion taken a prior course in ML/AI | 0.423 | 0.518 | |
| Proportion international student | 0.308 | 0.468 | |
| Proportion first generation student | 0.115 | 0.137 | |
| Proportion student with loans | 0.385 | 0.317 | |
| Proportion < 1 year ago | 0.154 | 0.252 | |
| Proportion 1-3 years ago | 0.538 | 0.583 | |
| Proportion 4-5 years ago | 0.269 | 0.129 | |
| Proportion 6-9 years ago | 0.038 | 0.000 | |
| Short-term intentional persistence | 0.840 (0.374) | 0.898 (0.304) | |
| Long-term intentional persistence | 0.731 (0.452) | 0.746 (0.437) | |
| Non-technical self-assessment | 4.308 (0.679) | 3.640 (0.747) | ** |
| Expertise Confidence | 3.077 (0.688) | 2.891 (0.780) | |
| Career-Fit Confidence | 2.731 (0.919) | 2.522 (0.776) | |
| Social Belonging Confidence | 2.654 (0.745) | 2.625 (0.807) | |
| Social Benefit Interest | 4.615 (0.496) | 4.346 (0.754) | |
| Toxicity of environment | 4.615 (0.983) | 4.254 (0.888) | ** |

| Variable | Not a visible minority (N=26) Mean (standard deviation) | Did not select "Not a visible minority" (N=139) Mean (standard deviation) | $\chi^2$ sig. test |
|---|---|---|---|
| Importance of high salary | 3.846 (0.925) | 4.111 (0.760) | |
| Competitive participation | 3.961 (1.148) | 3.044 (1.154) | ** |
| Influence of ML/AI popularity | 3.423 (1.172) | 3.706 (0.952) | |

*p < .05; **p < .01; ***p < .001 (two-tailed test).
Note that Fisher's Exact Test was used for any Chi-Square test where at least one tabulated frequency was less than 5 [99]

Appendix D: Bivariate Tests by international student status

Table 9: Bivariate Tests by international student status

| Variable | International student (N=73) Mean (standard deviation) | Not an international student (N=92) Mean (standard deviation) | $\chi^2$ sig. test |
|---|---|---|---|
| Proportion women | 0.274 | 0.380 | |
| Proportion man | 0.699 | 0.533 | * |
| | | | |
| Proportion Chinese | 0.466 | 0.380 | |
| Proportion South Asian | 0.123 | 0.228 | |
| Proportion not a visible minority | 0.110 | 0.196 | |
| | | | |
| Proportion 3rd Year | 0.178 | 0.446 | *** |
| Proportion 4th Year | 0.137 | 0.283 | * |
| Proportion 5th Year | 0.000 | 0.033 | |
| Proportion graduate studies | 0.685 | 0.228 | *** |
| | | | |
| Proportion Mechanical and Industrial Engineering major | 0.384 | 0.457 | |
| Proportion Electrical and Computer Engineering or Computer Science major | 0.260 | 0.239 | |
| Proportion ML specialist/minor/certificate | 0.521 | 0.620 | |
| Proportion taken a prior course in ML/AI | 0.534 | 0.478 | |
| | | | |
| Proportion first generation student | 0.164 | 0.109 | |
| | | | |
| Proportion student with loans | 0.192 | 0.435 | ** |
| | | | |
| Proportion < 1 year ago | 0.329 | 0.163 | * |
| Proportion 1-3 years ago | 0.507 | 0.630 | |
| Proportion 4-5 years ago | 0.123 | 0.174 | |
| Proportion 6-9 years ago | 0.000 | 0.011 | |
| | | | |
| Short-term intentional persistence | 0.908 (0.292) | 0.874 (0.334) | |
| Long-term intentional persistence | 0.836 (0.373) | 0.670 (0.473) | * |
| | | | |
| Non-technical self-assessment | 3.639 (0.756) | 3.833 (0.783) | |
| | | | |
| Expertise Confidence | 2.877 (0.781) | 2.956 (0.759) | |
| Career-Fit Confidence | 2.674 (0.708) | 2.462 (0.860) | |

| Variable | International student (N=73) Mean (standard deviation) | Not an international student (N=92) Mean (standard deviation) | $\chi^2$ sig. test |
|---|---|---|---|
| Social Belonging Confidence | 2.722 (0.755) | 2.556 (0.823) | |
| Social Benefit Interest | 4.472 (0.530)_ | 4.322 (0.846) | * |
| Toxicity of environment | 4.151 (0.953) | 4.440 (0.859) | |
| Importance of high salary | 4.194 (0.781) | 3.966 (0.790) | |
| Competitive participation | 3.292 (1.168) | 3.111 (1.222) | |
| Influence of ML/AI popularity | 3.708 (0.985) | 3.622 (1.001) | |

*$p < .05$; **$p < .01$; ***$p < .001$ (two-tailed test).
Note that Fisher's Exact Test was used for any Chi-Square test where at least one tabulated frequency was less than 5 [99]

Appendix E: Bivariate Tests by student loan status

Table 10: Bivariate Tests by student loan status

| Variable | Students with loans (N=54) Mean (Standard deviation) | Students without loans (N=111) Mean (Standard deviation) | $\chi^2$ sig. test |
|---|---|---|---|
| Proportion women | 0.204 | 0.396 | * |
| Proportion man | 0.778 | 0.523 | ** |
| | | | |
| Proportion Chinese | 0.296 | 0.477 | * |
| Proportion South Asian | 0.352 | 0.099 | *** |
| Proportion not a visible minority | 0.185 | 0.144 | |
| | | | |
| Proportion 3rd Year | 0.370 | 0.306 | |
| Proportion 4th Year | 0.241 | 0.207 | |
| Proportion 5th Year | 0.019 | 0.018 | |
| Proportion graduate studies | 0.370 | 0.459 | |
| | | | |
| Proportion Mechanical and Industrial Engineering major | 0.481 | 0.396 | |
| Proportion Electrical and Computer Engineering or Computer Science major | 0.241 | 0.252 | |
| Proportion ML specialist/minor/certificate | 0.556 | 0.586 | |
| Proportion taken a prior course in ML/AI | 0.500 | 0.505 | |
| | | | |
| Proportion first generation student | 0.167 | 0.117 | |
| | | | |
| Proportion international student | 0.259 | 0.532 | ** |
| | | | |
| Proportion < 1 year ago | 0.296 | 0.207 | |
| Proportion 1-3 years ago | 0.500 | 0.613 | |
| Proportion 4-5 years ago | 0.185 | 0.135 | |
| Proportion 6-9 years ago | 0.019 | 0.000 | |
| | | | |
| Short-term intentional persistence | 0.824 (0.385) | 0.921 (0.271) | |
| Long-term intentional persistence | 0.704 (0.461) | 0.764 (0.427) | |
| | | | |
| Non-technical self-assessment | 3.907 (0.734) | 3.667 (0.785) | |
| | | | |
| Expertise Confidence | 3.037 (0.751) | 2.864 (0.772) | |
| Career-Fit Confidence | 2.593 (0.922) | 2.536 (0.738) | |
| Social Belonging Confidence | 2.722 | 2.583 | |

|  | | |
|---|---|---|
|  | (0.811) | (0.787) |
| Social Benefit Interest | 4.315 | 4.426 |
|  | (0.843) | (0.659) |
| Toxicity of environment | 4.463 | 4.236 |
|  | (0.926) | (0.898) |
| Importance of high salary | 4.074 | 4.065 |
|  | (0.773) | (0.804) |
| Competitive participation | 3.444 | 3.065 |
|  | (0.461) | (1.162) |
| Influence of ML/AI popularity | 3.630 | 3.676 |
|  | (1.104) | (0.936) |

*$p < .05$; **$p < .01$; ***$p < .001$ (two-tailed test).
Note that Fisher's Exact Test was used for any Chi-Square test where at least one tabulated frequency was less than 5 [99]

Appendix F: Career Identity by student major. Percentages represent percentage of the row total.

| Major | "If I remain in ML/AI, I will identify as…." | | | | | |
| --- | --- | --- | --- | --- | --- | --- |
| | Computer Scientist | Data Scientist/ Data Analyst or Business Intelligence Professional | Engineer | More than one option | Other | Total |
| Electrical and Computer Engineering or Computer Science | N = 3 (7.32%) | N = 6 (14.63%) | N = 30 (73.17%) | N = 1 (2.44%) | N = 1 (2.44%) | N = 41 (100%) |
| Mechanical and Industrial Engineering | N=1 (1.49%) | N=28 (41.79%) | N=34 (50.75%) | N=2 (2.99%) | N=2 (2.99%) | N=67 (100%) |
| Other | N=2 (3.92%) | N=22 (43.14%) | N=23 (45.10%) | N=1 (1.96%) | N=3 (5.88%) | N=51 |
| Total | N = 6 (3.77%) | N = 56 (35.22%) | N = 87 (54.72%) | N = 6 (3.77%) | N = 4 (2.52%) | N = 159 |